\documentclass[twocolumn]{aastex62}
\pdfoutput=1 %for arXiv submission
\usepackage{epsfig}
\usepackage{amsmath,amstext}
\usepackage[T1]{fontenc}
\usepackage{footmisc}
\usepackage{natbib}
\usepackage[figure,figure*]{hypcap}

\newcommand{\Sp}{{\it Spitzer\/}}
\newcommand{\Sg}{{Sgr~A*}}

\shorttitle{\Sg\ Remarkable Flares}
\shortauthors{Fazio et al.}
\accepted{July 17, 2018 to the Astrophysical Journal}
\begin{document}

\title{Multiwavelength Light Curves of Two Remarkable Sagittarius A* Flares}

\author{G. G. Fazio}
\affiliation{Harvard-Smithsonian Center for Astrophysics, 60 Garden St., MS-65, Cambridge, MA 02138, USA} 
\author{J. L. Hora}
\affiliation{Harvard-Smithsonian Center for Astrophysics, 60 Garden St., MS-65, Cambridge, MA 02138, USA} 
\author{G. Witzel}
\affiliation{University of California - Los Angeles, Los Angeles, CA 90095, USA}
\author{S. P. Willner}
\affiliation{Harvard-Smithsonian Center for Astrophysics, 60 Garden St., MS-65, Cambridge, MA 02138, USA}
\author{M. L. N. Ashby}
\affiliation{Harvard-Smithsonian Center for Astrophysics, 60 Garden St., MS-65, Cambridge, MA 02138, USA}
\author{F. Baganoff}
\affiliation{Massachusetts Institute of Technology, 77 Massachusetts Ave, 37-555, Cambridge, MA 02139, USA}
\author{E. Becklin}
\affiliation{University of California - Los Angeles, Los Angeles, CA 90095, USA}
\author{S. Carey}
\affiliation{California Institute of Technology, MS 314-6, Pasadena, CA 91125, USA}
\author{D. Haggard}
\affiliation{McGill University, 3600 University Street, Montreal, QC, H3A 2T8, CANADA}
\author{C. Gammie}
\affiliation{University of Illinois - Urbana-Champaign, 1002 West Green Street, Urbana, IL 61801, USA}
\author{A. Ghez}
\affiliation{University of California - Los Angeles, Los Angeles, CA 90095, USA}
\author{M. A. Gurwell}
\affiliation{Harvard-Smithsonian Center for Astrophysics, 60 Garden St., MS-65, Cambridge, MA 02138, USA}
\author{J. Ingalls}
\affiliation{California Institute of Technology, MS 314-6, Pasadena, CA 91125, USA}
\author{D. Marrone}
\affiliation{University of Arizona, 933 North Cherry Avenue, Tucson, AZ 85721, USA}
\author{M. R. Morris}
\affiliation{University of California - Los Angeles, Los Angeles, CA 90095, USA}
\author{H. A. Smith}
\affiliation{Harvard-Smithsonian Center for Astrophysics, 60 Garden St., MS-65, Cambridge, MA 02138, USA}
\correspondingauthor{Giovanni G. Fazio}
\email{gfazio@cfa.harvard.edu}

\begin{abstract}
\Sg, the supermassive black hole (SMBH) at the center of our Milky Way Galaxy, is known
to be a variable source of X-ray, near-infrared (NIR), and submillimeter (submm) radiation and
therefore a prime candidate to study the electromagnetic radiation generated by mass accretion
flow onto a black hole and/or a related jet.  Disentangling the power source and emission
mechanisms of this variability is a central challenge to our understanding of accretion flows
around SMBHs.  Simultaneous multiwavelength observations of the flux variations and their time correlations can play an important role in obtaining a better understanding of possible
emission mechanisms and their origin.  This paper presents observations of two flares that both
apparently violate the previously established patterns in the relative timing of submm/NIR/X-ray
flares from \Sg. One of these events provides the first evidence of coeval structure between NIR and submm flux increases, while the second event is the first example of the sequence of submm/X-ray/NIR flux increases all occurring within $\sim$1~hr.  Each of these two events appears to upend assumptions that have been the basis of some analytic models of flaring in \Sg.  However, it cannot be ruled out that these events, even though unusual, were just coincidental.  These observations demonstrate that we do not fully understand the origin of the multiwavelength variability of \Sg, and show that there is a continued and important need for long-term, coordinated, and precise multiwavelength observations of \Sg\ to characterize the full range of variability behavior.

\end{abstract}

\tighten

\section{Introduction}

Sagittarius A* (\Sg), the supermassive black hole (SMBH) at the center of our Milky Way galaxy, is 100 times closer than any other SMBH and is therefore a prime candidate to study the electromagnetic radiation generated by mass accretion flow onto a black hole and/or a related jet.  SgrA* has been targeted for decades in attempts to learn about SMBH physics from its variability at many wavelengths.  Rapid fluctuations have been detected at soft and hard X-ray bands (Baganoff et al.\ 2001; Nowak et al.\ 2012; Neilson et al.\ 2013, 2015; Barri\`ere et al.\ 2014; Ponti et al.\ 2015) and at near-infrared (NIR) wavelengths (Genzel et al.\ 2003; Ghez et al.\ 2004; Hornstein et al.\ 2007; Witzel et al.\ 2012, 2018; Hora et al.\ 2014), where the extinction to the Galactic center is relatively low.  Slower and lower-amplitude variability has been seen at submillimeter (submm), millimeter (mm), and radio wavelengths (Mauerhan et al.\ 2005; Macquart \& Bower 2006; Marrone et al.\ 2008; Yusef-Zadeh et al.\ 2006b; Brinkerink et al.\ 2015).  Disentangling the power source and emission mechanisms of the variability is a central challenge to our understanding of accretion flows around SMBHs at the cores of normal galaxies.   The chief barrier to progress is the absence of a sufficient sample of simultaneous, multiwavelength variability measurements, their correlations being key to discriminating among emission models (Neilson et al.\ 2015; Dibi et al.\ 2016; Connors et al.\ 2017).  

X-ray variability of \Sg\ has been observed by Chandra, XMM-Newton, and Swift at energies of 2 -- 8 keV and by NuStar at 2 -- 79 keV.  X-ray flares are observed at the rate of $\sim$1.1 per day and typically last $\sim$0.8~hr (Neilson et al.\ 2015; Ponti et al.\ 2015).  There is often temporal substructure within X-ray flares (Baganoff et al 2001, 2003; Goldwurm et al.\ 2003; Porquet et al.\ 2003, 2008; Eckart et al.\ 2004, 2006a; Belanger et al.\ 2005; Trap et al.\ 2011; Nowak et al.\ 2012; Neilsen et al.\ 2013; 2015; Degenaar et al.\ 2013; Barri\`ere et al.\ 2014) even on timescales as short as $\sim$100~s (Nowak et al.\ 2012; Barri\`ere et al.\ 2014; Neilson et al.\ 2015).  The fast variability indicates that the flares come from regions smaller than $\sim$10 light minutes (<15 Schwarzschild radii).  Flare amplitudes range from a few to 10 times the quiescent flux with rarer flares extending to factors of up to 400 times quiescence (Porquet et al.\ 2003; Dodds-Eden et al.\ 2011; Nowak et al.\ 2012; Ponti et al.\ 2017).  The quiescent flux ($L_{(2 - 10\ {\rm keV})}\sim2\times10^{33} $erg s$^{-1}$) is spatially extended ($\sim 1''$) and extremely faint (Baganoff et al.\ 2003; Wang et al.\ 2013), while the flaring source is $\leq 1''$ in size and bright.

At NIR wavelengths, as observed by VLT and Keck (1.6 -- 3.8~\micron), there is a continuously variable flux that exceeds the measurement noise level (0.033 mJy for VLT and 0.017 mJy for Keck) about 60\% of the time for the VLT observations and 90\% of the time for Keck observations (Meyer et al.\ 2014; Witzel et al.\ 2012).  With \Sp, emission at 4.5~\micron\ was detected above the noise level $\sim$34\% of the time for 1 minute averaging and >60\% of the time for 6.4 minute averaging (Hora et al.\ 2014; Witzel et al.\ 2018).  Observed flux densities at 2.2~\micron\ ranged from 0 to $\sim$8 mJy (Witzel et al.\ 2018).  Although the NIR light curve superficially resembles a quiescent state interrupted by brief flares, that appearance is misleading.  Witzel et al.\ (2018) showed that the underlying probability density function of NIR flux densities is continuous and well fit by a log-normal distribution, and the time-dependence is a random walk process.  The apparent flares are merely the highest levels within the continuous flux-density distribution, and there is continuous variability below the typical noise level.  This gives the impression of discrete outbursts, so referring to NIR peaks as ``flares'' is an apt description of the observed behavior at typical S/N level, but the source has no separate flaring state.  Brighter flux density levels ($\sim$5 -- 10 mJy) occur about 4 -- 5 times per day (Witzel et al.\ 2012; Hora et al.\ 2014).  These events have a duration of $\sim$0.5 to 2 hours with substantial temporal substructure including variations on timescales of 47~s (Dodds-Eden et al.\ 2009).  As with X-rays, the temporal substructure implies emission from very compact regions.  The NIR emission is strongly polarized, suggesting its origin is synchrotron radiation (Eckart et al. 2006b).  The spectral index $\alpha$\  $(S_\nu \propto \nu^\alpha)$ measured between 1.6 and 3.7~\micron\ during bright output intervals is 0.6~$\pm$~0.2 (Hornstein et al.\ 2007), which suggests that the NIR photons are optically thin synchrotron emission from electrons with a power-law energy spectrum.  The spectral index is redder during fainter intervals (Witzel et al.\ 2018).

At millimeter/submm wavelengths, as observed by the Submillimeter Array (SMA; 890~\micron\ and 1.3~mm), the Atacama Pathfinder Experiment telescope (APEX; 870~\micron), and the Caltech Submillimeter Observatory (CSO; 450 and 850~\micron), there is a relatively continuous emission at flux densities of $\sim$3 Jy at 850~\micron\ and 3.5 Jy at 1300~\micron.  This is the highest luminosity observed at any wavelength.  In addition to the continuous emission, there are $\sim$1.2 flaring events per day on time scales of hours to days with amplitudes $\sim$25\% or greater of the continuous emission (Eckart et al.\ 2006a; Yusef-Zadeh et al.\ 2008; Marrone et al.\ 2006; 2008).  However, our observations (2004 -- 2007) of submm emission using the SMA and ALMA produced a rate of $\sim$4.6/day.  The submm continuous emission level of \Sg\ varies and, as a result, it is difficult to determine what is a flare event.  Linear polarization has also been observed in submm flare emission, increasing from 9\% to 17\% as the flare progresses through its temporal peak (Marrone et al.\ 2006).

Herschel/SPIRE monitored \Sg\ at far-infrared wavelengths (250, 350, and 500~\micron) and observed highly significant variations with temporal structure that was highly correlated across these wavelengths, but the variability amplitude was a maximum of $\sim$0.4 Jy, which is $\leq$20\% of the 450~\micron\ flux density from \Sg\ (Stone et al.\ 2016; Marrone et al.\ 2006).  Recently von Fellenberg et al.\ (2018), during one observation obtained with Herschel/PACS, measured simultaneous variation of \Sg\ at 160 and 100~\micron.  The observed peak emission from the median flux at 160~\micron\ was (0.27~$\pm$~0.07) Jy and at 100~\micron\ was (0.16~$\pm$~0.10) Jy.

Following the discovery of NIR and X-ray variability, coordinated observing campaigns over a range of wavelengths were initiated.  The first flare of \Sg\ to be observed simultaneously at NIR and X-rays occurred in 2004 (Eckart et al.\ 2004).  However, it has proven very difficult to obtain a sufficient number of simultaneous flares at X-ray and NIR wavelengths to determine their relationship (Eckart et al.\ 2006b, 2008; 2012; Yusef-Zadeh et al.\ 2006a, 2012; Hornstein et al.\ 2007; Marrone et al.\ 2008; Dodds-Eden et al.\ 2009; Ponti et al.\ 2017).  In a total of $\sim$60 hours in blocks of 3 to 7 hours of simultaneous NIR and X-ray observations using VLT/Keck and Chandra, there were 5 X-ray flares observed with corresponding NIR emission.  In 3 of the events, peak flux occurred nearly simultaneously at both wavelengths (Figure 1).  Strong X-ray flares always showed a coincident NIR maximum, but there were numerous NIR maxima with no X-ray counterpart (Eckart et al.\ 2004; Hornstein et al.\ 2007; Figure 1).  Yusef-Zadeh et al.\ (2012) and Ponti et al.\ (2017) found evidence that X-ray peak emission lagged behind the NIR peak emission with a time delay ranging from a few to tens of minutes.  The substructure and FWHM of coincident events differed at the two wavelengths, with the FWHM of the NIR emission wider than the X-ray emission. 

\begin{figure*}
\begin{center}
\includegraphics[width=0.7\textwidth]{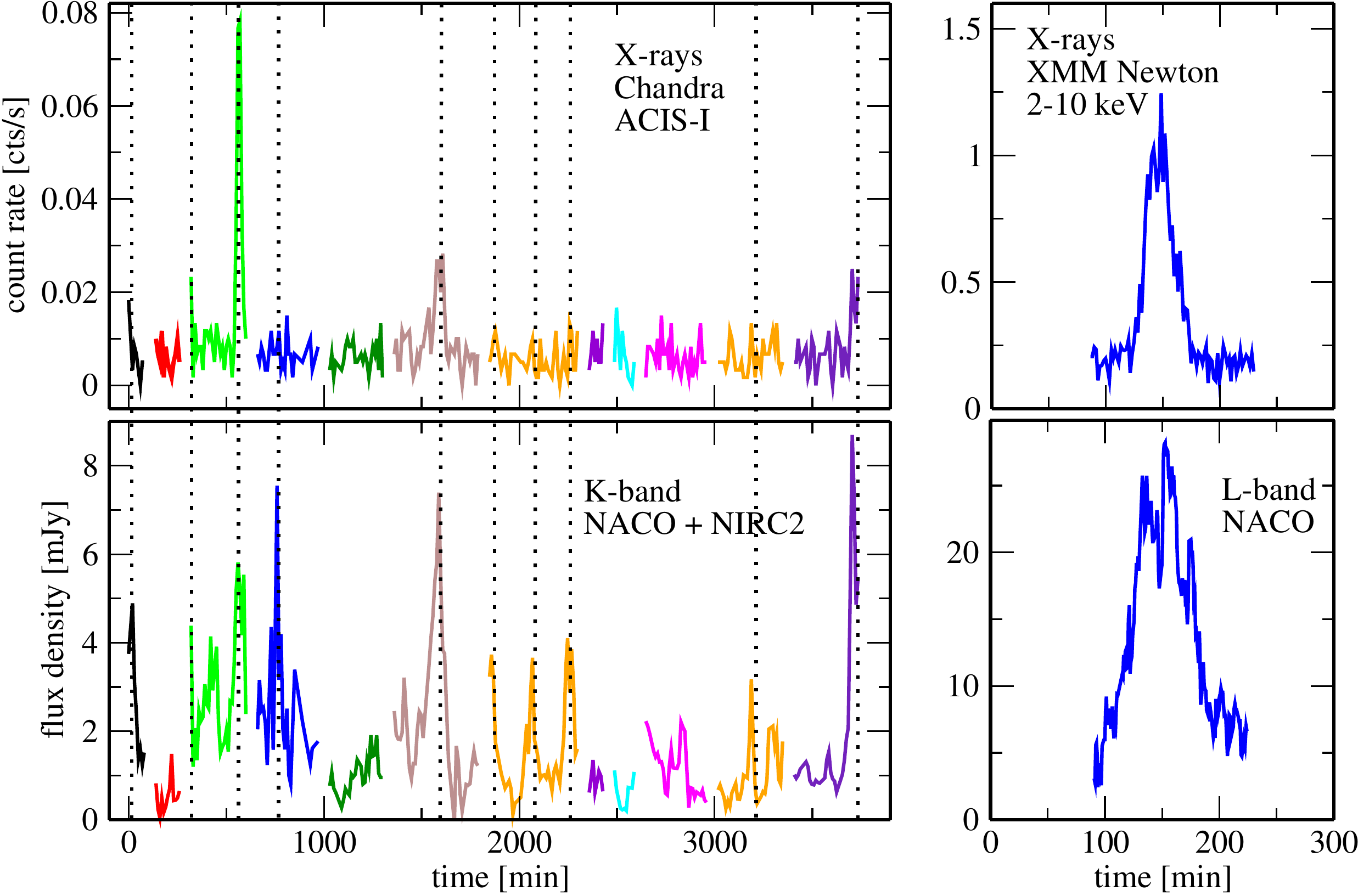}\label{fvsdiam}
\end{center}
\caption{Left: A total of $\sim$58~hrs of simultaneous observations with VLT/NACO and Keck/NIRC2 at 2.2~\micron\ and Chandra X-ray Observatory/ACIS-I at 2 -- 8 keV is displayed.  Observations were taken in separate 3 -- 7 hour intervals as shown by the colors but are merged here on a continuous time axis.  Vertical lines mark NIR peaks. The minimum observed flux density of 3.05 mJy was subtracted from the 2.2~\micron\ data before plotting. This graph was provided by Zhiyuan Li.  Right: The X-ray (2  --  10 keV; XMM-Newton) and the L-band (3.8~\micron; VLT/NACO) data from the very bright 2007 April 4 flare (Dodds-Eden et al.\ 2009).
}
\end{figure*}
Associating radio/submm flares with NIR and X-ray flares has been more difficult than establishing the NIR/X-ray connection.  The first observations of a flare of \Sg\ detected at submm, NIR, and X-ray wavelengths were reported by Marrone et al.\ (2008).  There are other examples of seemingly related events seen in the submm near the time of NIR and/or X-ray flares (Eckert et al.\ 2008).  However, given the multi-hour duration of typical submm brightening events, the short submm observing windows (nearly all less than 8 h), and the infrequency of these events, there is some chance that the submm events are only coincidentally related to the shorter-wavelength activity.  Using only the Keck data (Marrone et al.\ 2008), an apparent delay of 20 minutes was reported between the NIR and 1.3 mm peaks.  However, when the VLT NIR data were added, extending the light curve, a 160 minute delay was observed (Meyer et al.\ 2008).  Morris et al.\ (2012) summarized the reported time lags between NIR and millimeter/submm peaks for 7 events.  The delays ranged from 90 minutes to 200 minutes, with one possible exception, but on the average, the delay appeared to be $\sim$150 min.

The origin of the NIR and X-ray brightening fluctuations remains unknown, but the rapid modulation suggests the emission source most likely originates just outside the event horizon.  Study of the light curves may therefore provide insight into the structure and conditions in the inner accretion regions.  (See review by Morris et al.\ 2012.)  The near-simultaneity of the NIR and X-ray brightness fluctuations suggests a common origin, but the fact that not all NIR peaks are accompanied by X-ray flares suggests that either there are two physical origins for the NIR events or that the physical mechanism for the flares has two different observational manifestations, e.g., different flares arise from electron energy distributions having different high-energy cutoffs, so some have enough energy to emit X-rays and some don't.

Despite the past multiwavelength monitoring over several years, \Sg\ can still surprise us.  This paper presents observations of two flares that seem to violate the previous patterns in the relative timing of submm/NIR/X-ray flares from \Sg.  One of these flares provides the first evidence of near time coincidence between a NIR and submm outburst, while the second event is the first example of an X-ray flare followed by a submm outburst followed by a NIR peak, all within $\sim$1~hr.

This paper is organized as follows.  Section 2 describes the data.  The flares and their properties are shown in Section 3, and Section 4 offers possible interpretations for the origin of these events.

\section{Observations}\label{ObsRed}
Two epochs of simultaneous observations of \Sg\ are reported here.  The first was on 2014 June 17 -- 18 with the \Sp\ Space Telescope/IRAC at 4.5~\micron\ and the Submillimeter Array (SMA) at 395 GHz (890~\micron).  The second was on 2015 May 14 with the Keck telescope at 2.12~\micron, the SMA at 395 GHz, and the Chandra X-ray Observatory at 2 -- 8 keV. 

\subsection{\Sp/IRAC Observations}
All \Sp/IRAC (Fazio et al.\ 2004) observations used here were part of the \Sp\ PID 10060 program (PI: G. Fazio; Hora et al.\ 2014), which observed \Sg\ at 4.5~\micron\ during four epochs of $\sim$23.4 hours each.  IRAC was operated in subarray mode, which obtains 64 consecutive images (a ``frame set'') of a 32$\times$32 pixel region (1.21$''$ per pixel) near the corner of the 256$\times$256 pixel array.  The observations used a subarray frame time of 0.1~s for a duration of 6.4~s for each frame set.  The observing sequence after the first slew to the field included an initial mapping operation and then two successive 11.6-hour staring mode observations, each using the ``PCRS Peak-up'' to center \Sg\ on pixel (16,16) of the subarray.  The first stare began at 2014 June 17 18:57:17 UT (HMJD 56825.7907247). The details of the data acquisition are described by Hora et al.\ (2014) and the data reduction by Witzel et al.\ (2018), who also provided the observed light curves.  Because of the data reduction method, the zero point of the IRAC light curves is undetermined.  Tabulated flux densities are relative to the average value when \Sg\ was in a low-emission mode.  

%full lightcurve plots

\subsection{Keck observations}
Observations amounting to 144 minutes were conducted with the adaptive optics (AO)-supported near-infrared camera NIRC2 at the Keck II telescope of the W. M. Keck Observatory on Mauna Kea, Hawaii.  We used standard observation scripts for the central $10''\times10''$ of our Galaxy. The same scripts have been used to track the orbital motions of stars at the Galactic center (e.g., Ghez et al.\ 2005, Ghez et al.\ 2008, Meyer et al.\ 2014; Boehle et al.\ 2016).  The observation started at 2015-05-14 12:18:10.7 UT, and the last integration ended at 14:02:01.9 UT (MJD 57156.51262  --  57156.58474). 105 frames were taken in K-band ($\lambda$ = 2.12~\micron; $\Delta \lambda$ = 0.35~\micron) in a small offset dither pattern with a cadence of $\sim$15 seconds.  For these observations the laser guide star position was fixed in the center of the image, and USNO-A2.0 0600-28577051 (R = 13.7) was used as tip and tilt guide star.

The observing conditions were variable, and the laser launch telescope temporarily iced over, resulting in variable FWHM values of 70 mas to 110 mas and Strehl ratios consistently below 20\%.  However, we were able to make use of 100 frames for photometry on \Sg. Following the steps described by Witzel et al.\ (2012), we derived for each frame a point spread function (PSF) using the PSF-fitting package StarFinder (Diolaiti et al.\ 2000) and deconvolved the innermost $4'' \times 4''$ of the frame.  We then used aperture photometry (circular apertures with radius of 30 mas) to determine the brightness of 13 calibrators, \Sg, and 6 background apertures. We used the calibrator magnitudes listed by Witzel et al.\ (2012), and we subtracted the measured average sky brightness. As a result we obtained a light curve for \Sg\ with an accuracy of 5 --  7\% in relative photometry.

\subsection{The millimeter/submm observations}
On 2014 June 18 the SMA observed \Sg\ from UT 7.4 to 13.5 hours (start time $\sim$HMJD 56826.3083) at 342.971 GHz (875~\micron) with eight antennas operational in the compact configuration (baseline lengths of 14.9 -- 99.4~k$\lambda$).  The total continuum bandwidth was $\sim$8~GHz.  The SMA was operated in a dual-receiver polarization track with double sideband observations using sideband separation implemented in the correlator.  During the observations the precipitable water vapor (pwv) ranged from 1.2 to 1.6~mm.  The continuum visibility was calculated by averaging the two same-sense polarization signals.  The gain calibration source was QSO J1733--1304  (=NRAO 530), and Neptune was used as the flux calibration source.  The final flux density measurement was determined by vector-averaging the measured visibility data for instantaneous baselines greater that 40~k$\lambda$ to avoid including large-scale emission structure around \Sg.

On 2015 May 14 the SMA again observed \Sg\ from UT 9.3 to 16.1 hours (MJD 2457156.89024) at 226.881 GHz (1.32 mm) with six antennas operational in the extended configuration (baseline lengths of 12.1 -- 129.5~k$\lambda$).  However, after UT 13.5 hours correlator issues affected the data resulting in reduced temporal coverage.  The total continuum bandwidth was $\sim$8 GHz.  The SMA was operated in a single receiver polarization track with double sideband observations using sideband separation implemented in the correlator.  At the startup of the observations the pwv was 1.8 mm, dropped within 1.5 hours to 0.9 mm, and then dropped more slowly to 0.7 mm.  The continuum visibility and gain calibration were as the 2014 June 18 observations but with Calisto and Titan as the flux calibrators.  The final flux density measurement again used only baselines >40 k$\lambda$.

\begin{figure*}
\begin{center}
\includegraphics[width=0.7\textwidth]{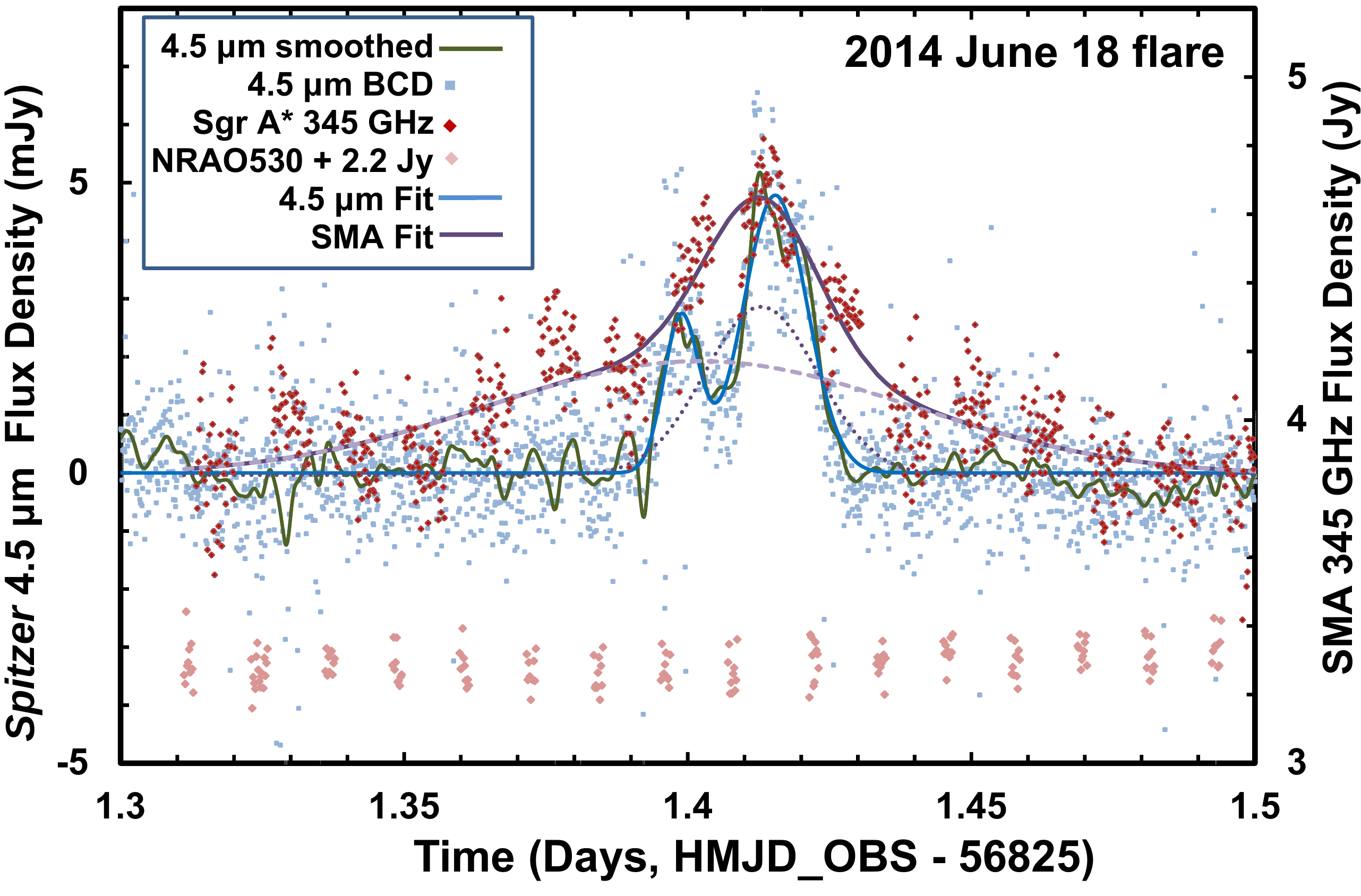}\label{fvsdiam}
\end{center}
\caption{The 2014 June 18 joint observations of a double-peaked flare from \Sg\ by \Sp/IRAC at 4.5~\micron\ (blue dots and green smoothed line, scale on the left ordinate) and the SMA at 890~\micron\ (red dots, scale on the right ordinate).  The SMA calibrator (NRAO 530) flux density (shown in light red at the bottom) is $\sim$1 Jy (a constant of 2.2 Jy was added to place the data on the right ordinate scale).  The blue smoothed line is the two-Gaussian curve fit to the 4.5~\micron\ data.  Dotted and dashed purple lines show two Gaussian curves fitting the SMA submm data, and the solid purple line shows their sum.
}
\end{figure*}

\begin{deluxetable*}{cccccc}
\tablecaption{\Sg\ Flare of 2014 June 18 UT}
\tablecolumns{6}
\tablewidth{0pt}
\tablehead{
\colhead{Telescope} & \colhead{Wavelength} &
\colhead{Time (HMJD)} &
\colhead{Amplitude} &
\colhead{Width} &
\colhead{Delay from} \\
\colhead{} & \colhead{} & \colhead{} & \colhead{above} & \colhead{(min)} &
\colhead{summed submm}\\
& & & \colhead{baseline} & \colhead{(FWHM)} & \colhead{peak (min)}
}
\startdata
\Sp/IRAC	 & 4.5~\micron	& 56826.398857	&2.7 mJy	&10.5	&-19.6\\
SMA	 & 890~\micron&	 56826.402082	&0.33 Jy	&126.8	&-15.0\\
SMA	& 890~\micron	& 56826.412940	&0.49 Jy	&33.8	&+0.7\\
\Sp/IRAC	 & 4.5~\micron	  & 56826.415429&	4.8 mJy	&19.1&	+4.3\\
\enddata
\end{deluxetable*}
\subsection{X-ray observations}
\Sg\ was observed during Chandra's Cycle 16 on 2015 May 14 beginning at 08:45:44 UTC (MJD 57156.36509) for a total of 22.74 ks (ObsID 16966). The data were acquired using the ACIS-S3 chip in FAINT mode with a 1/8 sub-array centered on the radio position of \Sg: 17:45:40.0409,  --29:00:28.118 (J2000.0; Reid \& Brunthaler 2004).  The small sub-array was adopted to mitigate photon pileup in bright flares from \Sg\ and to avoid contamination from nearby X-ray binaries; it has the ancillary advantage of achieving a frame time of $\sim$0.4~s (vs. the standard Chandra frame time of 3.2 s).  Chandra data reduction and analysis were performed with standard CIAO v.4.8 tools (Fruscione et al.\ 2006)\footnote{Detailed information about the Chandra Interactive Analysis of Observations (CIAO) software
is available at http://cxc.harvard.edu/ciao/} and calibration database v4.7.2. We reprocessed the level 2 events file with the Chandra repro script to apply up-to-date calibrations and then extracted the 2 -- 8 keV light curve from a circular region with a radius of 1\farcs25 (2.5 pixels) centered on \Sg.  The small extraction region and energy filter help isolate \Sg's flare emission and minimize contamination from diffuse X-ray background emission (e.g., Baganoff et al.\ 2001, Nowak et al.\ 2012, Neilsen et al.\ 2013). 

\section{Results}\label{periods}

\subsection{2014 June 18 UT}

We observed a significant double-peaked NIR flare on 2014 June 18 at $\sim$09:45 UT (HMJD 56826.41209) that occurred during simultaneous observations with the SMA (890~\micron).  Light curves are shown in Figure 2.  There was no Chandra coverage of this event, and NuSTAR observations were heavily contaminated by the transient outburst of AX J1745.6-2901, so it is not known whether the flare produced any X-ray emission.  (AX J1745.6-2901 is located 2\farcs4 from \Sg\ and is not a problem for Chandra with its 0\farcs5 angular resolution.)

The NIR light curve can be fit with two Gaussian peaks separated by 23.9 minutes. The submm light curve is well fit by the sum of two Gaussian functions with much broader FWHMs than the NIR flares.  The fit parameters and the delay time with respect to the summed submm peak are given in Table 1.  The baseline flux density of the submm data is 3.84 Jy.

\begin{figure*}
\begin{center}
\includegraphics[width=0.7\textwidth]{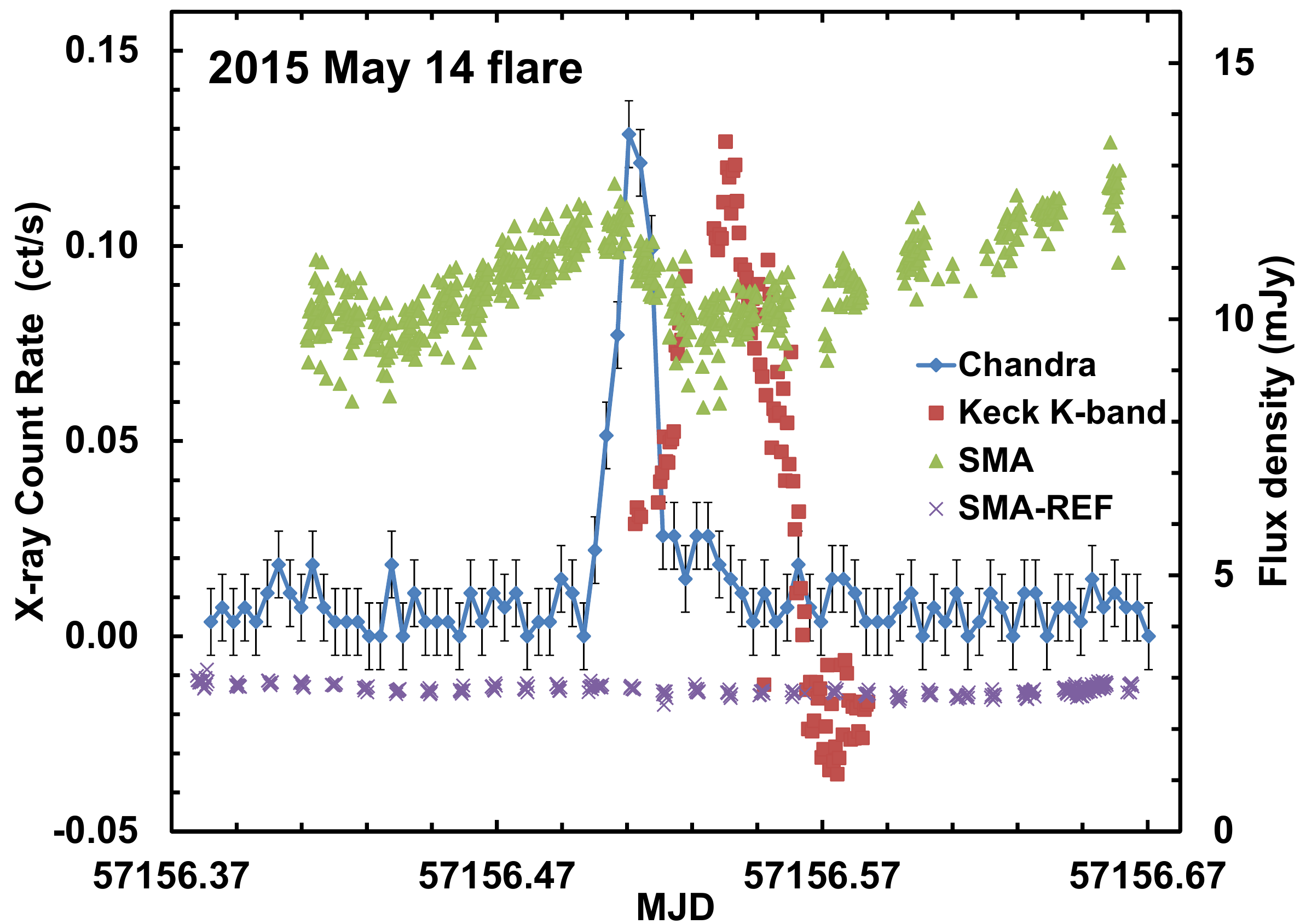}\label{fvsdiam}
\end{center}
\caption{2015 May 14 observations of a single-peaked flare from \Sg.  Red squares show the Keck 2.2~\micron\ data (scale on right ordinate), green triangles the 890~\micron\ SMA data (scale is 500x the right ordinate, i.e., peak flux density is $\sim$6 Jy), and blue points the Chandra 2 -- 8 keV data (left ordinate). The X-ray light curve is summed over 300~s bins, and Poisson error bars on the X-ray count rate are shown by black error bars.  The SMA calibrator (NRAO 530) flux density ($\sim$1.4 Jy), scaled the same as for \Sg, is shown by purple crosses.}
\end{figure*}

\subsection{2015 May 14 UT}

Simultaneous observations of \Sg\ were achieved by Keck K-band (2.12~\micron), Chandra X-ray Observatory, and the SMA (890~\micron) on 2015 May 14 during a flare at $\sim$11:50 UT (MJD 57156.49311258).  In this unusual event a single-peaked flare was coincident at X-ray and submm wavelengths, but the NIR peak was delayed by 65.5 minutes (Figure 3).  Light curves at each wavelength can be fit by a baseline plus single Gaussian, and the results are summarized in Table 2.  The NIR outburst of \Sg\ reached $\sim$10.4 mJy, more than half the flux density of the star S0-2.  This flare is unprecedented in several aspects: the submm peak precedes the X-ray peak by 27.7 minutes, and the NIR peak is delayed from the peak of the submm emission by 65.5 minutes.  This has never been seen in previous events.   The observation time before and after the flare is very limited, and it is possible that the peaks at different wavelengths may be associated with other flares that either preceded or followed the observed flare.  In particular, the Keck observations cannot rule out a NIR peak occurring before or simultaneously with the X-ray and submm peaks.

\section{Discussion}

The relative timing of flux increases observed at X-ray, NIR, and submm wavelengths from \Sg\ can be an important tool to constrain the physical processes of how plasma behaves close to the event horizon of a SMBH.  

A number of simplified phenomenological models have been proposed to explain the observed time delays in \Sg.  In one model, the emission arises from adiabatically expanding, relativistic ($\sim$0.1c) electron clouds (plasma blobs) that emit through the synchrotron self-absorbed mechanism.  The plasma blobs are initially opaque at submm wavelengths, but as they expand, they become cooler and optically thin, reaching an emission maximum at progressively longer wavelengths with time (van der Laan et al.\ 1966; Yusef-Zadeh et al.\ 2006b, 2008, 2009; Eckart et al.\ 2006a, 2006b, 2008, 2009; Marrone et al.\ 2006, 2008).  The plasma blob model also predicts that the X-ray and NIR emissions should occur nearly simultaneously because they are both optically thin throughout the expansion and come from the same source, with the X-rays arising from optically-thin synchrotron self-Compton (SSC), inverse Compton processes, or synchrotron emission (Markoff et al.\ 2001; Yuan et al.\ 2004; Eckart et al.\ 2006a, b, 2009, 2012; Liu et al.\ 2004, 2006; Yuan et al.\ 2004; Yusef-Zadeh et al.\ 2006a, 2008, 2012; Dodds-Eden et al.\ 2009).  The SSC/expanding blob model has been used to explain the submm flare time delay of $\sim$150 minutes from the X-ray/NIR outburst (Yusef-Zadeh et al.\ 2006b, 2008, 2009; Marrone et al.\ 2008; Eckart et al.\ 2008, 2012; Wardle 2011; Morris et al.\ 2012; Dexter et al.\ 2014).  

A jet model with shock heating of non-thermal electrons at the jet base has been proposed by Falcke \& Markoff (2000) and Markoff et al.\ (2001).  Falcke, Markoff \& Bower (2009) have proposed a pressure driven, mildly relativistic jet-like outflow also based on the SSC/expanding blob model.  This model has been used to explain the time delay between submm, millimeter, and radio outbursts, with the radio to submm time lag being $\sim$65 minutes between 7 and 0.86 mm wavelength (Yusef-Zadeh et al.\ 2006b, 2008; Bower et al.\ 2015; Brinkerink et al.\ 2015; Subroweit et al.\ 2017).  

Other proposed models that produce some of the characteristics of the multiwavelength observations include transient orbiting hot spots, which are the product of shocks and magnetic reconnection events within the accretion flow (Broderick \& Loeb 2006), and shock heating in a tilted accretion disk Dexter \& Fragile (2013).  Dodds-Eden et al.\ (2009, 2010) and Dibi et al.\ (2014) have proposed a model based on a cooling break in the synchrotron radiation by non-thermal electrons in magnetized (5 -- 30 gauss) plasma during an accretion flow to explain X-ray and NIR simultaneous events.

\begin{deluxetable*}{cccccc}
\tablecaption{\Sg\ Flare of 2015 May 14 UT}
\tablecolumns{6}
\tablewidth{0pt}
\tablehead{
\colhead{Telescope} & \colhead{Wavelength} &
\colhead{Time (HMJD)} &
\colhead{Amplitude} &
\colhead{Width} &
\colhead{Delay from} \\
\colhead{} & \colhead{} & \colhead{} & \colhead{above} & \colhead{(min)} &
\colhead{summed submm}\\
& & & \colhead{baseline} & \colhead{(FWHM)} &\colhead{peak (min)}}
\startdata
SMA &890~\micron & 57156.49311258	&0.809  Jy	&30.5&	0\\
Chandra	&4 keV&	57156.51233664	&0.1234 ct/s&	7.9&	27.7\\
Keck	&2.1~\micron&	57156.53859682	&10.36 mJy&	22.1 &	65.5\\
\enddata
\end{deluxetable*}
Magnetohydrodynamic (MHD) models of accretion flow have been proposed (Goldston et al.\ 2005), and most recently Moscibrodzka et al.\ (2014), Chan et al.\ (2015), and Ball et al.\ (2016) have created general relativistic magnetohydrodynamic (GRMHD) numerical simulation models to explain the flux increases across the electromagnetic spectrum and their relative time delays.  The Chan et al.\ (2015) standard and normal (SANE) models reproduce the short-timescale millimeter and NIR variability with amplitudes and power spectra that are comparable to those observed.  Flares in these models arise from strong-field gravitational lensing near the horizon of emission from short-lived, hot, magnetically dominated flux tubes.   The SANE models also produce correlated variability only between 1.3 mm and NIR curves with small ($\leq$1~hr) time lags, and no correlated variability between radio, 1.3 mm, and X-ray wavelengths. This model also provides a natural explanation for the observed NIR flares with no X-ray counterpart and favors NIR emission originating in the accretion disk rather than a jet.  In contrast, the Chan et al.\ (2015) magnetically arrested disk (MAD) models produce variability only at lower flux levels, and the radio, 1.3 mm, and NIR light curves are highly correlated, with marginal evidence of the NIR leading the 1.3 mm light curves and the radio light curves lagging the 1.3 mm by $\sim$1~hr.  A similar level of correlation exists between the inner X-ray and 1.3 mm light curves, with the former leading the latter by $\sim$2~hr.  The predicted sequence is flux increasing in the X-rays first, then the NIR, then the 1.3 mm, and then the radio flux.  In the MAD models the conjecture is that magnetic reconnection is responsible for the variability.  Neither the SANE nor the MAD models produce large X-ray flares, but Ball et al.\ (2016) incorporated large X-ray variability by ``localizing non-thermal electrons to highly magnetized regions, where particles are likely to be accelerated via magnetic reconnection.  The proximity of these high-field regions to the event horizon forms a natural connection between NIR and X-ray and accounts for the rapid timescales associated with the X-ray flares.''

The 2014 June 18 flare (Figure 2), with the smoothed peaks of the NIR and submm fluxes within $\sim$4 minutes of each other, and the fact that the submm flux rises earlier that the NIR flux, differs from previous results that the submm flares follow concurrent NIR and X-ray outbursts with time delays of $\sim$150 minutes.  This flare sequence seems to rule out the SSC/expanding blob model and/or the jet model as its origin. The near-simultaneous NIR and submm peaks may be consistent with the Chan et al.\ (2015) GRMHD models invoking strong lensing.  Connecting the X-ray/NIR models to the submm/radio models has always been difficult.  However, there is the chance, for this flare, that the submm event is only coincidentally related to the NIR event.   The random coincidence rate (e.g., Evans 1955) for NIR/submm events, assuming for the NIR 4 events/day of 1 hour duration and for the submm 1.2 events/day of 2 hour duration, is 0.60/day.  If we assume a submm event rate of 4.6/day, then the rate is 2.3/day.  With 32 hours of NIR/submm simultaneous observations from 2014 to 2017, it is possible this event is a random coincidence, even though the peak emissions of the two curves coincide within $\sim$4 minutes.  

For the 2014 June 18 flare, the submm flux density declines from the peak by 50\% in just  23~$\pm$~5 minutes.  Although each submm data cluster may have a full amplitude total spread of $\sim$0.4~Jy, each cluster has 30 to 40 points, and the average amplitude is therefore known with an uncertainty of $\sim$0.1~Jy.  The flare amplitude (peak to baseline; 4.65 to 3.84 Jy) is 0.81~Jy.  Therefore the S/N is $\sim$8, and the peak is significant.  The submm calibrator has a noise level of $\pm$0.05~Jy.   Marrone et al.\ (2008) observed a submm decay approximated by an exponential decay time of $\sim$2~hr, similar to that observed by Eckart et al.\ (2006b).  This time is much shorter than the cooling time scale due to synchrotron losses in a $\sim$20~G magnetic field.   That cooling time is $\sim$25 minutes in the NIR, and the cooling time for the submm would be $\sim$20 times slower.

The 2015 May 14 flare (Figure 3) is unique in that the submm flux density peaks first, followed $\sim$28 minutes later by the X-ray peak, with the NIR peak emission occurring $\sim$66 minutes after the submm peak.  No model we are aware of explains this sequence.  However, the NIR observations began only at the peak of the X-ray flare, at which time the NIR flux was already elevated over the noise level.  Because NIR events are often multi-peaked, a second, smaller NIR peak could have been associated with the X-ray flare with the subsequent, larger NIR peak being one with no X-ray association.  In addition, \Sg\ is always varying at mm wavelength about a mean level of $\sim$3 Jy.  Again, there is the chance that in the case of this flare, the X-ray, NIR, and submm flux increases are only coincidently related to each other.  The random coincidence rate for X-ray/submm events, assuming for the X-ray 1.1 events/day of 1 hour duration and for the submm 1.2 events/day of 2 hour duration, is 0.15/day.  If we assume a submm event rate of 4.6/day, then the random coincidence rate is 0.59/day.

The temporal structure of the two flare events shown in this paper implies that current theoretical models cannot explain the origin of the multiwavelength variability of \Sg.  There is a continued and important need for long-term, coordinated, and precise multiwavelength observations of \Sg\ in order to characterize the full range of flare behaviors.  The X-ray spectral index and its possible correlation with the characteristics of the NIR or submm activity (peak intensity, time lags, peak duration, rise or fall times) might hold a clue to both the X-ray emission mechanism and to the underlying cause of the variability.  Future coordinated monitoring should endeavor to go all the way to cm-wave radio in order to determine the wavelength dependence of the phase lags, as the results in this paper show that our present view of the physical processes at work and their wavelength dependences are far from being understood.  There is also a need for more detailed and more varied models of strongly sub-Eddington accretion onto supermassive black holes.

\acknowledgements

This work is based on observations made with the \Sp\ Space Telescope, which is operated by the Jet Propulsion Laboratory, California Institute of Technology under a contract with NASA.  We thank the staff of the \Sp\ Science Center for their help in planning and executing these demanding observations.  The W. M. Keck Observatory is operated as a scientific partnership among the California Institute of Technology, the University of California, and the National Aeronautics and Space Administration.  The Observatory was made possible by the generous financial support of the W. M. Keck Foundation.  The Submillimeter Array is a joint project between the Smithsonian Astrophysical Observatory and the Academia Sinica Institute of Astronomy and Astrophysics and is funded by the Smithsonian Institution and the Academia Sinica.  The scientific results reported in this article are based in part on observations made by the Chandra X-ray Observatory.  CFG was supported by NSF 13-33612 and NSF 17-16327.  HS acknowledges partial support of NASA Grant NNX14AJ61G.  Support for the UCLA participants for this work was provided by NSF grant AST-14-12615.  Support for CfA participants was provided by NASA Grant 80NSSC18K0416.  We wish to thank Zhiyuan Li for providing Figure 1.

%\software{IRAF, mopex \citep{makovoz06}, IRACproc \citep{schuster06}}

\facilities{\Sp/IRAC}%,GTC
\tighten
\begin{center}
REFERENCES
\end{center}
\vskip 1pt

\hangindent=0.7cm 

Baganoff, F. K., et al.\ 2001, Nature, 413, 45

Baganoff, F. K., et al.\ 2003, ApJ, 591, 891

Ball, D. et al.\ 2016, ApJ, 826, 77

Barri\`ere, N. M. et al.\ 2014, ApJ, 786, 46

Belanger, G et al.\ 2005, ApJ, 635, 1095

Boehle, A., Ghez, A. M., \& et al.\ 2016, ApJ, 830, 17

Bower, G. C. et al.\ 2015, ApJ, 802, 69

Brinkerink, C. D. et al.\ 2015, A\&A, 576, 41

Broderick, A. E. and Loeb, A. 2006, JPhCS, 54, 448

Chan, C. et al.\ 2015, ApJ, 812, 103

Connors, R. M., et al.\ 2017, MNRAS, 466, 4121

Degenaar, N. et al.\ 2013, ApJ, 769, 155

Dexter, J. and Fragile, P. C., 2014, MNRAS, 432, 2252

Dibi, S. et al.\ 2014, MNRAS, 441, 1005

Dibi, S. et al.\ 2016, MNRAS, 461,552

\hangindent=0.7cm 
Diolaiti, E., Bendinelli, O., Bonaccini, D., et al.\ 2000, Proc. SPIE Vol. 4007, p. 879-888

Dodds-Eden, K. et al., 2009, ApJ, 698, 676

Dodds-Eden, K. et al., 2010, ApJ, 725, 450

Dodds-Eden, K. et al., 2011, ApJ, 728, 37

Eckart, A. et al.\ 2004, A\&A, 427, 1

Eckart, A. et al.\ 2006a, A\&A, 450, 535 

Eckart, A. et al.\ 2006b, A\&A, 455, 1

Eckart, A. et al.\ 2008, A\&A, 492, 337

Eckart, A. et al.\ 2009, A\&A, 500, 935

Eckart, A. et al.\ 2012, A\&A, 537, A52

\hangindent=0.7cm 
Evans, R. 1955, The Atomic Nucleus, p. 791 (New York: McGraw-Hill Book Company, Inc.)

Falcke, H. \& Markoff, S. 2000, A\&A, 362, 113

\hangindent=0.7cm 
Falcke, H., Markoff, S. \& Bower, G. C.  2009, A\&A, 496, 77

Fazio, G. et al.\ 2004, ApJ, 154, 10

\hangindent=0.7cm 
Fruscione, A., McDowell, J. C., Allen, G. E., et al.\ 2006, Proc. SPIE, 6270, 62701

Genzel, R. et al.\ 2003, Nature, 425, 934

Ghez, A., et al.\ 2004, ApJ, 601, L159

\hangindent=0.7cm 
Ghez, A. M., Hornstein, S. D., Lu, J. R., et al.\ 2005, ApJ, 635, 1087 

\hangindent=0.7cm 
Ghez, A. M., Salim, S., Weinberg, N. N., et al.\ 2008, ApJ, 689, 1044

Goldsston, J. E., et al.\ 2005, ApJ, 621, 785

Goldwurm, A., et al.\ 2003, ApJ, 584, 751

Hora, J. L. et al.\ 2014, ApJ, 793, 120

Hornstein et al.\ 2007, ApJ, 667

Liu, S., et al.\ 2004, ApJ, 611, L101

Liu, S., et al.\ 2006, ApJ, 647, 1099

Macquart, J.-P. \& Bower, G. C. 2006, ApJ, 646, 111

Markoff, S., et al.\ 2001, A\&A, 379, L13

Mauerhan, J. C., et al.\ 2005, ApJ, 623, 25

Marrone, D. P. et al.\ 2006, ApJ, 640,308

Marrone, D. P. et al.\ 2008, ApJ, 682, 373

Meyer, L. et al.\ 2008, ApJ, 688, L17

\hangindent=0.7cm 
Meyer, L., Witzel, G., Longstaff, F. A., \& Ghez, A. M. 2014, ApJ, 791, 24

Morris, M. R. et al.\ 2012, RAA, 12, 995

Moscibrodzka, M., et al.\ 2014, A\&A, 570, A7

Neilsen, J., et al.\ 2013, ApJ, 774, 42

Neilsen, J., et al.\ 2015, ApJ, 799, 199

Nowak, M. A., et al.\ 2012, ApJ, 759, 95

Ponti, G., et al.\ 2015, MNRAS, 454, 1525

Ponti, G., et al.\ 2017, MNRAS, 468, 2447

Porquet, D. et al.\ 2003, A\&A, 407, L17

Porquet, D. et al.\ 2008, A\&A, 488, 549

\hangindent=0.7cm 
Reid, M. J. \& Brunthaler, A. 2004, ApJ, 616, 872

Stone, J. M., et al.\ 2016, ApJ, 825, 32S

Subroweit, M., et al.\ 2017, A\&A, 601, A80 

Trap, G., et al.\ 2011, ApJ, 528, A140

\hangindent=0.7cm 
von Fellenberg, S. D., et al.\ 2018, arXiv:1806.07395v1

Wang, Q. D. et al 2013, Science, 341, 981

van der Laan, H. 1966, Nature, 211, 1131

Wardle, M. 2011, ASP Conf Ser. Vol. 439, 450

\hangindent=0.7cm 
Witzel, G., Eckart, A., Bremer, M., et al.\ 2012, ApJS, 203, 18

\hangindent=0.7cm 
Witzel, G., et al.  2018, accepted for publication in ApJ, arXiv:1806.00479

Yuan, F., et al.\ 2004, ApJ, 606, 894

Yuan, F., et al.\ 2014, ARA\&A, 52, 529

Yusef-Zadeh, F. et al.\ 2006a, ApJ, 644, 198

Yusef-Zadeh, F. et al.\ 2006b, ApJ, 650, 189, 

Yusef-Zadeh, F. et al.\ 2008, ApJ, 682, 361

Yusef-Zadeh, F. et al.\ 2009, ApJ, 706, 348

Yusef-Zadeh, F. et al.\ 2012, AJ, 144, 1

\end{document}